\documentclass[%
 reprint,
 amsmath,amssymb,
 aps,
 pre,
]{revtex4-1}

\usepackage{graphicx}
\usepackage{dcolumn}
\usepackage{bm}
\usepackage{url}

\usepackage{amsmath}
\newcommand{\mean}[1]{\langle #1 \rangle}

\usepackage{appendix}
\usepackage[bf]{subfigure}
\usepackage[utf8]{inputenc}
\newcommand{\inst}[1]{$^#1$}

\begin{document}

\preprint{APS/123-QED}

\title{Epidemic spreading with awareness and different timescales in multiplex networks}

\author{Paulo Cesar Ventura da Silva\inst{1}}
\author{F\'{a}tima Vel\'{a}squez-Rojas\inst{8}} 
\author{Colm Connaughton $^{3, 4}$} 
\author{Federico Vazquez $^{8,9}$}
\author{Yamir Moreno $^{5,6, 7}$}
\author{Francisco A. Rodrigues $^{2, 3,4}$}
\email{francisco@icmc.usp.br}
\affiliation{
  \inst{1} Instituto de F\'{i}sica de  S\~{a}o Carlos, Universidade de S\~{a}o Paulo, S\~{a}o Carlos, SP, Brazil.\\         
  \inst{2} Instituto de Ci\^{e}ncias Matem\'{a}ticas e de Computa\c{c}\~{a}o, Universidade de S\~{a}o Paulo, S\~{a}o Carlos, SP, Brazil.\\
  \inst{3} Mathematics Institute, University of Warwick, Gibbet Hill Road, Coventry CV4 7AL, UK.\\
  \inst{4} Centre for Complexity Science, University of Warwick, Coventry CV4 7AL, UK.\\
  \inst{5} Institute for Biocomputation and Physics of Complex Systems (BIFI), University of Zaragoza, 50018 Zaragoza, Spain\\
    \inst{6}Department of Theoretical Physics, University of Zaragoza, 50018 Zaragoza, Spain\\
    \inst{7} Complex Networks and Systems Lagrange Lab, Institute for Scientific Interchange, Turin, Italy\\
    \inst{8} Instituto de F\'{i}sica de L\'{i}quidos y Sistemas Biol\'{o}gicos (UNLP-CONICET), 1900 La Plata, Argentina\\
    \inst{9} Instituto de C\'{a}lculo, FCEN, Universidad de Buenos Aires and CONICET, Buenos Aires, Argentina
}

\begin{abstract}
One of the major issues in theoretical modeling of epidemic spreading is the development of methods to control the transmission of an infectious agent. Human behavior plays a fundamental role in the spreading dynamics and can be used to stop a disease from spreading or to reduce its burden, as individuals aware of the presence of a disease can take measures to reduce their exposure to contagion. In this paper, we propose a mathematical model for the spread of diseases with awareness in complex networks. Unlike previous models, the information is propagated following a generalized Maki-Thompson rumor model. Flexibility on the timescale between information and disease spreading is also included. We verify that the velocity characterizing the diffusion of information awareness greatly influences the disease prevalence. We also show that a reduction in the fraction of unaware individuals does not always imply a decrease of the prevalence, as the relative timescale between disease and awareness spreading plays a crucial role in the systems' dynamics. This result is shown to be independent of the network topology. We finally calculate the epidemic threshold of our model, and show that it does not depend on the relative timescale. Our results provide a new view on how information influence disease spreading and can be used for the development of more efficient methods for disease control.
\end{abstract}


\maketitle

\section{Introduction}

Mathematical and computational studies of epidemic models have proven to be very important for understanding real-world disease dynamics~\cite{satorras2015,DeArruda2018}. Currently, one of the main motivations behind epidemic modeling is the development of methods and models that allow to control the transmission of an infectious agent~\cite{satorras2015, Wang016Vaccine}. These methods include the optimization of more traditional strategies to control an outbreak, such as vaccination~\cite{Wang016Vaccine, Dorso017} or or a quarantine mechanism based on adaptive connections~\cite{Lagorio011,vazquez2016rescue}, but also novel approaches that take into account more accurately human behavioral responses
.

Modeling the influence of human behavior in disease spreading is an intense research topic \cite{funk2010modelling,manfredi2013modeling,Meloni2011,wang2015coupled,Funk2015NineChallenges}. In particular, individual prevention methods can considerably reduce the overall incidence of a disease, but the acknowledgement of the methods and the decision to adopt them depend on behavioral factors. The latter have been modeled using opinion dynamics \cite{salathe2008,eames2009networks,Velasquez_2017,Pires_2018}, game-theoretical approaches\cite{Bauch2004,Fu2011,Reluga2010,chen2018suppressing_social}, spreading processes \cite{Funk6872,arenas_2013,arenas_2014,asymmetrically_wang_2014,Wang2016Supressing,guo2015cascade}, risk perception \cite{risk_perception_single_layer,wu_2012,risk_perception_multiplex,poletti2012risk,moinet2018risk} and other approaches \cite{wang2015coupled}.

The risk perception approach considers that individuals become aware of an epidemics by noticing the presence of infected individuals in their neighboring contacts. Bagnoli and others \cite{risk_perception_single_layer} showed that individual protection triggered by risk perception can stop an epidemics from spreading in several network topologies, including a moment-diverging scale-free (provided that the perception response is nonlinear). On the other hand, awareness by spreading phenomenon models the word-of-mouth propagation of an information about the epidemics. Funk and others \cite{Funk6872} showed that, for an SIR epidemics, the spreading of awareness through individual contacts could avoid an epidemic outbreak, whereas a global awareness could only reduce the outbreak size, but not stop it. Wu and others \cite{wu_2012} also studied the influence of global awareness, risk perception and contact-spreading awareness in an SIS epidemics, showing that the local (but not the global) awareness could raise the epidemic threshold. These works highlighted the importance of local information for controlling epidemics.

In more recent works, Granell et al. \cite{arenas_2013,arenas_2014} and Wang et al. \cite{asymmetrically_wang_2014,Wang2016Supressing} also studied spreading awareness in SIS and SIR epidemic models, respectively. In these works, it was shown that there can be an information outbreak either triggered by itself or triggered by an epidemic outbreak, thus depicting an ``information without disease'' stage on the models' phase diagrams. Due to this state, Granell and collaborators \cite{arenas_2013} show that there is a \emph{metacritical point} for the epidemic threshold values, and that a global source of information (regarded as a mass media campaign) can eliminate this metacritical point, as it causes the awareness to be always present \cite{arenas_2014}. Besides, on the SIR framework, Wang and collaborators \cite{Wang2016Supressing} show that there is an optimal information transmission rate that minimizes the disease spreading. 

Most of these works, however, consider that the epidemics and the awareness propagate and vanish at the same timescale. This is a limitation, as many real-world phenomena do not occur at the same rates. For instance, the HIV infection and renewing cycle has a typical timescale of many years, whereas the information and awareness about HIV can be spread and forgotten several times during this period. Moreover, important discoveries were obtained by flexibilizing relative timescales between simultaneous processes. For example, Oliveira and Dickman \cite{Oliveira2017} determined that the competition between two biological species may be won by that with a slower birth/death rate under certain environmental circumstances. G\'omez and collaborators \cite{Gomez2013diffusion} showed that the diffusion on a two-layer network can be possibly faster than on each individual layer, if the diffusion time-scales inside and between the layers are different. It may, therefore, be of great value to consider that epidemic spreading and the propagating awareness about it occur at different timescales.

Moreover, the models for epidemics and awareness proposed until the present use simple models for the spreading of awareness. Bagnoli \cite{risk_perception_single_layer} considers a \emph{fading} awareness, which looses its quality as it propagates and eventually extinguishes. Granell \cite{arenas_2013,arenas_2014} applies a simple SIS epidemic model for the awareness, in which spreaders stop propagating the information by forgetting it (analogous to the disease healing). These approaches capture the essential phenomenology, but may be not as accurate in reproducing the real-world behavior, in which people may know the information but loose the interest in propagating it. This can be considered using \emph{rumor models} \cite{daley_kendall_1965,moreno2004,trpevski2010}, in which a compartment of individuals called \emph{stiflers} is used to represent people who does not want to propagate the information that they have.

For the modeling of the interaction between epidemics and awareness, as well as other interacting processes, multilayer networks are a very useful tool \cite{Kivela014,boccaletti2014structure,Cozzo2013,Cozzo2018,PhysRevLett.96.138701,bianconi2013statistical,mucha2010community,demeester1999resilience}. In particular, multilayers in which all layers have the same number of nodes (often called multiplex networks) can be used to model interacting phenomena that do not share the same contact structure, so that each layer encodes the contacts associated to the respective dynamics. Most of the works on disease-behavior interaction that we mentioned previously on this paper use multiplexes for their modeling.


Here we explore a model in which two processes coexist: the spreading of a disease and the dissemination of awareness of the disease. Our model includes two new ingredients. First, we increase the complexity of previous models with respect to the dissemination of information by considering the dynamics of the Maki-Thompson \cite{maki_thompson_1974} rumor model $-$ instead of using the traditional dynamics of disease spreading~\cite{arenas_2013,arenas_2014}. Second, we introduce a parameter that allows to control the relative timescales between the disease and rumor propagation processes. Results for scale-free networks show that the rumor dynamics can indeed reduce the epidemic prevalence. However, if we couple the characteristic time for awareness diffusion with the state of individuals, namely, by considering that infected individuals take more time to inform about its own infection, a counterintuitive behavior is revealed: the prevalence increases with the rate at which individuals become aware, despite the fact that fraction of unaware individuals decreases. The latter mechanism is important, as there are many diseases to which a similar behavior can be associated $-$ e.g., HIV transmission, where HIV-positive patients are often reluctant to voluntarily notify their sexual partners~\cite{Njozing011}. In what follows, we present the model as well as some analytical insights and results from numerical simulations. We round off the paper by discussing our findings and presenting possible applications to the modeling of real diseases.
 
\section{The model}
Our model considers the propagation of a disease in a population, simultaneously to the spreading of information about it, by which individuals become aware of the disease and of prevention methods, reducing their contagion probabilities. These two processes run in a double-layer multiplex network: one layer for the disease spreading and another one for the information awareness to hold the disease. As of the definition of a multiplex~\cite{Kivela014}, each layer has the same number of nodes, and there is a one-to-one link between the nodes in different layers. In this sense, we identify each pair of linked nodes from each layer as the same ``individual'';  the only difference from one layer to the other one lies in the structure of connections inside the layers. The links on the ``epidemic layer'' represent contacts that can possibly transmit the disease, whereas links on the ``informational layer'' represent pairs of individuals that share information with each other, like in social online networks. 

\subsection{Baseline model}

The model for the epidemic spreading adopted here is a reactive SIS (susceptible-infected-susceptible) compartmental model~\cite{satorras2015,DeArruda2018} in which, at each time step $\Delta t=1$ of the dynamics, each infected (I) node tries to transmit the disease to each of its susceptible (S) neighbors on the epidemic layer with probability $\beta$, and then tries to recover with probability $\mu$. 

For the spreading of information awareness to prevent the transmission we use a cyclic Maki-Thomson rumor model in complex networks~\cite{Nekovee07}, which we call UARU (unaware-aware-stifler-unaware). Notice that the latter R here is used for the stifler compartment to avoid confusion with the susceptible (S) state in the SIS model. A stifler is an informed node who does not propagate the information anymore. When an aware (spreader) node contacts an unaware (ignorant) neighbor in the informational layer, it tries to pass the rumor about the disease. If the contacted neighbor, however, is an aware or stifler node, the node that makes the contact becomes stifler. A stifler individual can also forget  the  information  about  the  disease  transmission, becoming ignorant about the disease transmission again.  We again use a discrete time approach~\cite{gomez_2010} by considering a reactive formulation in which, at each time step $\Delta t=1$, each aware (A) node first tries to inform each of its unaware (U) neighbors with probability $\gamma$, and then becomes a stifler (R) with probability $\sigma$.  Besides, each stifler node becomes ignorant (U) with probability $\alpha$.

By combining the epidemic and the informational states of each node, we can describe the overall state of each individual. In our model, we have six different states, i.e., SU (susceptible and unaware), SA (susceptible and aware), SR (susceptible and stifler), IU (infected and unaware), IA (infected and aware) and IR (infected and stifler). Using these overall states, we define the interaction between the epidemics and awareness by adding two new features. First, a susceptible node that is informed (aware or stifler) will reduce its contagion probability by a factor $\Gamma$ (with $0 \leq \Gamma < 1$) for each contact, meaning that it will get the disease from each of its infected neighbors with probability $\Gamma \beta$ (less than $\beta$). Such a feature represents the adoption of prevention methods against the disease. Second, an additional transition called \emph{self-awareness} is considered: if not informed by a neighbor, an infected-unaware (IU) node can, during the same time step, become aware with probability $\kappa$, by knowing its own condition. This process simulates the case in which an infected subject recognizes the symptoms of the disease and becomes aware of the infection.

\begin{figure}
    \includegraphics[width=0.40\textwidth]{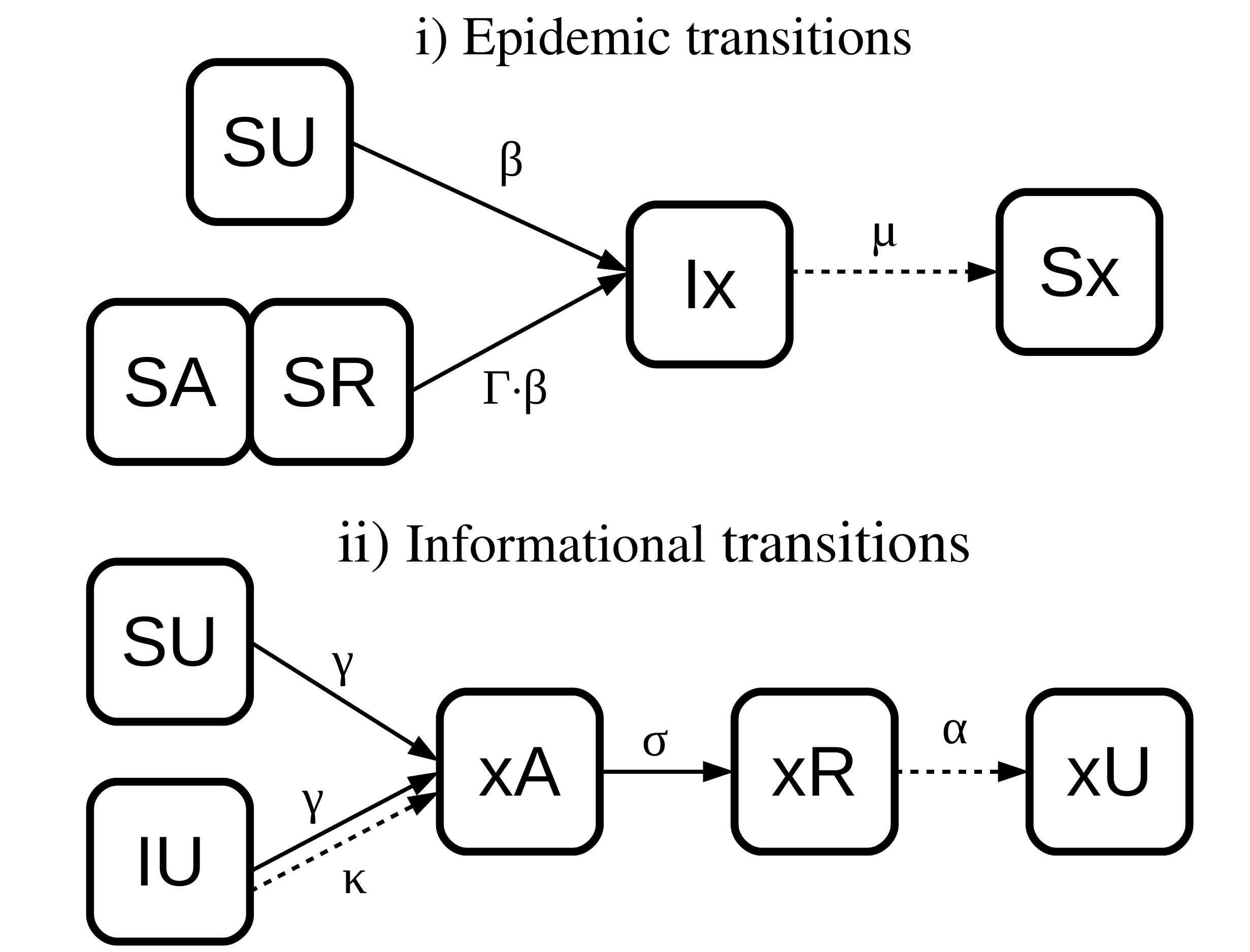}
    \caption{Schematic illustration showing the states of the nodes in the network and the associated transition probabilities between states, indicated by the Greek letters.}
    \label{fig:sisuaru_transitions}
\end{figure}


The following reaction equations - representing respectively the (\ref{eq:epid_1}) infection of an unaware susceptible, (\ref{eq:epid_2}) infection of an aware susceptible, (\ref{eq:epid_3}) infection of a stifler susceptible and (\ref{eq:epid_4}) healing of an infected node - describe all possible epidemic transitions (where $x$ is used to represent an arbitrary informational state):

\begin{eqnarray}
    \label{eq:epid_1} SU + Ix &\overset{\beta}{\longrightarrow}& IU + Ix, \\
    \label{eq:epid_2} SA + Ix &\overset{\Gamma\beta}{\longrightarrow}& IA + Ix, \\
    \label{eq:epid_3} SR + Ix &\overset{\Gamma\beta}{\longrightarrow}& IR + Ix, \\
    \label{eq:epid_4} Ix &\overset{\mu}{\longrightarrow}& Sx. 
\end{eqnarray}

The informational transitions - respectively (\ref{eq:rumor_1}) information of an unaware node, (\ref{eq:rumor_2}) self-awareness of an infected unaware, (\ref{eq:rumor_3}) ``stifling'' of an aware node by contacting another aware node, (\ref{eq:rumor_4}) ``stifling'' of an aware via contact with a stifler and (\ref{eq:rumor_5}) forgetting of the information - are represented by these equations ($x$ and $y$ represent arbitrary epidemic states):

\begin{eqnarray}
    \label{eq:rumor_1} xU + yA &\overset{\gamma}{\longrightarrow}& xA + yA, \\
    \label{eq:rumor_2} IU &\overset{\kappa}{\longrightarrow}& IA, \\
    \label{eq:rumor_3} xA + yA &\overset{\sigma}{\longrightarrow}& xR + yA, \\
    \label{eq:rumor_4} xA + yR &\overset{\sigma}{\longrightarrow}& xR + yR, 
    \\  
    \label{eq:rumor_5} xR &\overset{\alpha}{\longrightarrow}& xU.
\end{eqnarray}

Figure \ref{fig:sisuaru_transitions} presents the possible transitions between the six states, grouped according to the epidemic and informational dynamics.

The timescale of the model is controlled according to a defined probability.  With probability $\pi$, only the rumor transitions (awareness, self-awareness, stifling and forgetting) can happen during the current time step. With the complementary probability $(1 - \pi)$, the epidemic transitions (infection and recovering) can occur. By setting the value of $\pi$, it is possible to emulate different timescales between the two processes. For instance, a value of $\pi$ close to $1.0$ means that the rumor propagates much faster than the infectious agent.

\subsection{Modified model}
\label{sec:modified_model}

Besides the baseline model that we have described, we propose a minor modification that can generate some unexpected behaviors. We extend the idea of self-awareness to stifler nodes, considering that a stifler, which is also infected by the disease, is less likely to forget the information. That is, a node who knows about its own infection does not inform other nodes and also impair the transmission of other nodes, creating additional stiflers around it. This behavior is approximately observed in the case of HIV transmission, in which some infected individuals knows about its own infection but do not voluntarily notify their sexual partners~\cite{Njozing011}, acting as infected-stiflers. 
We include this feature by reducing the probability that an infected-stifler node forgets the information by a factor of $(1 - \kappa)$ (so that the self-awareness parameter also reduces the rate at which infected-stiflers become infected-unaware). We refer to this version of the model as \emph{modified model}, whereas the version without this modification is referred to as \emph{baseline model}.

In the following section, we describe our results with both baseline and modified models.

\begin{figure}[!th]
    \centering
    \includegraphics[width=0.48\textwidth]{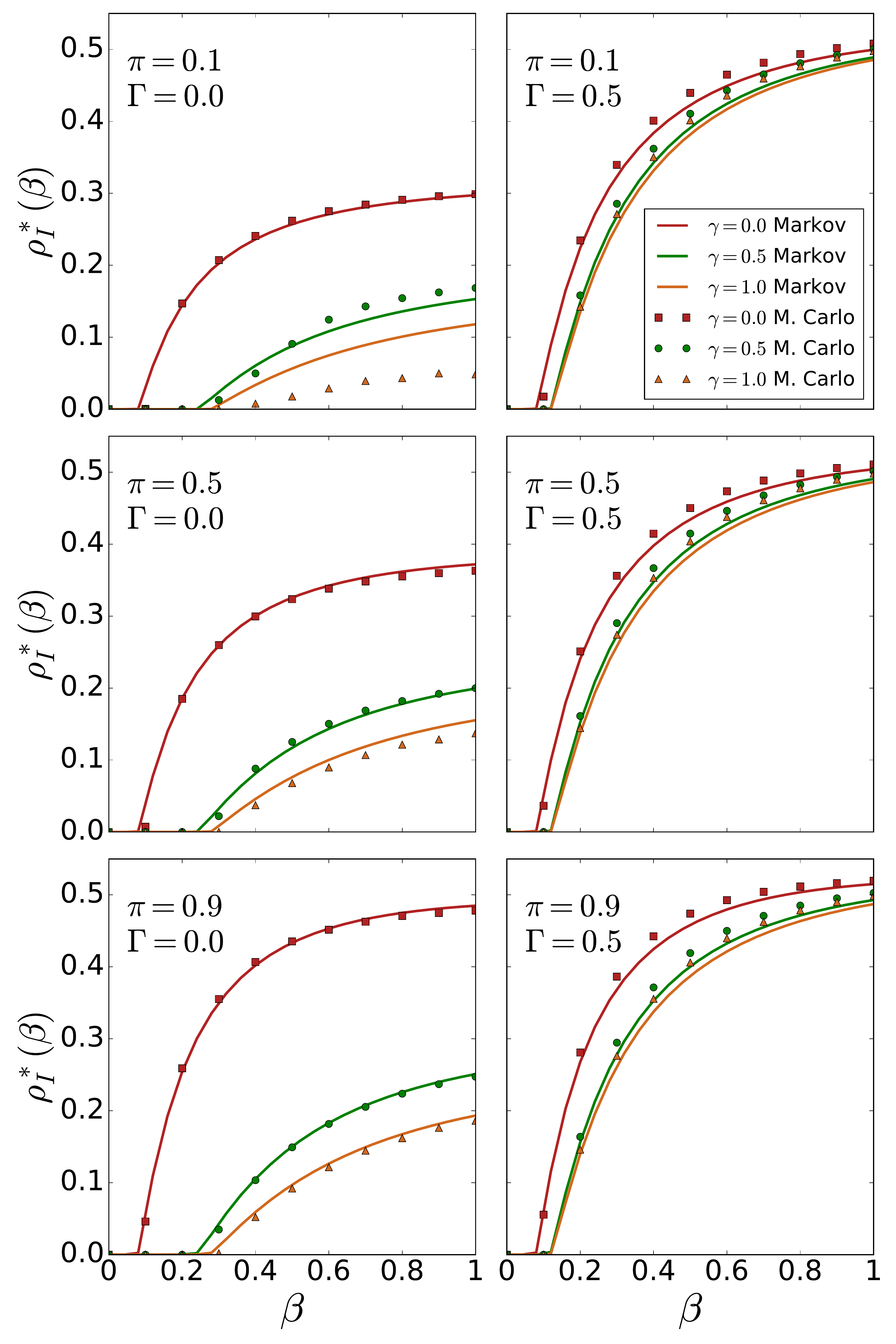}
    \caption{Stationary density of infected nodes $\rho_I^*$ (disease prevalence) as a function of the disease propagation probability ($\beta$), for different values of $\gamma$, $\Gamma$ and $\pi$, using the baseline model. For $\pi = 0.5$, the rumor spreading and the epidemic propagation have the same time scale, whereas for $\pi = 0.1$ ($0.9$) rumor events are slower (faster) than the events of the epidemic process. The solid lines are Markov chain calculations (see appendix), whereas symbols are results from Monte Carlo simulations. Other parameters of the model are set to: $\mu = 0.9$, $\alpha=0.6$, $\kappa = 0.5$ and $\sigma=0.6$.}
    \label{fig:rhoi_vs_beta}
\end{figure}

\section{Results}
We performed extensive Monte Carlo (MC) simulations of the dynamics described in the last section, where we considered a multiplex network composed by two layers with scale-free organization and $N=1000$ nodes each.  Each layer was generated independently by using the configuration model~\cite{Bender78} with power-law exponent $\gamma_{sf} \approx -2.5$ and minimum degree $k_{min} = 4$, with a resulting average degree $\langle k \rangle \approx 7.4$ in each layer. The node correspondence between the two layers is done at random, generating thus no relevant degree correlation for corresponding nodes in each layer. In order to further study the model, we also developed a Markov chain approach that consists of solving a set of fixed point equations that provide the stationary fractions of nodes in each state. The Markov chain method is described in the appendix at the end of this paper. 

In figure \ref{fig:rhoi_vs_beta}, the stationary density of infected nodes $\rho_I^*$ (prevalence) is plotted against the infection probability $\beta$, for different values of the parameters $\gamma$ (information spreading probability), $\Gamma$ (immunization factor for informed nodes) and $\pi$ (relative time scale). For this first analysis, we only used the baseline model. Symbols represent the results from MC simulations of the model, while solid lines correspond to the solution of the Markov chain approach.  
To calculate stationary densities by MC simulations, we run the dynamics for $T = 1200$ time steps, ignore the first $400$ time steps and calculate the average fraction of nodes in the desired state over the remaining $800$ steps. Each data point corresponds to an average over $10^3$ independent realizations of the dynamics. At the initial state of each realization, $20\%$ of the nodes are randomly chosen and assigned the state IA (infected-aware), whereas the remaining $80\%$ of nodes are set to the SU state (susceptible unaware). For the Markov chain calculations, initially each node begins the process with probability $p^i_{IA}(0)=0.2$ for the infected-aware state and $p^i_{SU}(0)=0.8$ for the susceptible-unaware state, with the remaining state probabilities being set to zero.

Analyzing figure \ref{fig:rhoi_vs_beta}, we can first check that the information about the disease helps in both reducing the prevalence and increasing the epidemic threshold, by comparing curves with different values for the information spreading probability $\gamma$. Moreover, the prevalence is decreased when the immunity provided by the awareness is total ($\Gamma = 0.0$) rather than partial ($\Gamma = 0.5$). However, the prevalence is also increased if the relative timescale $\pi$ is greater, i.e., when the transitions of the rumor process are faster than those of the epidemic process. This is an intriguing result as, intuitively, we expect that a faster informational process should be more efficient in preventing the disease spreading.  We believe that an insight into this counterintuitive effect can be obtained by studying simpler versions of the present model within a mean-field approach, which we left for future work.

\begin{figure}[!t]
\begin{center}
\includegraphics[width=0.4\textwidth]{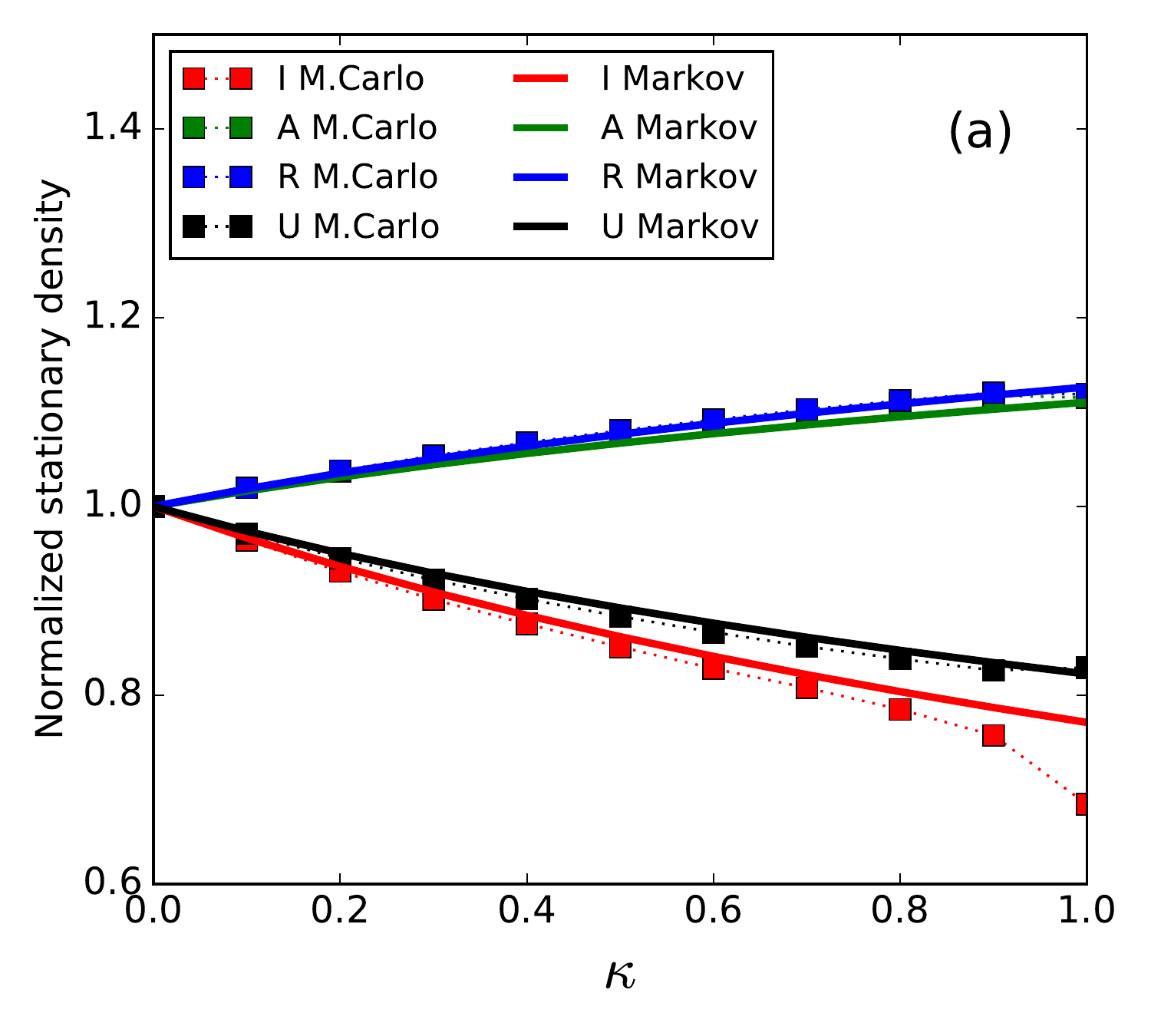}
\includegraphics[width=.4\textwidth]{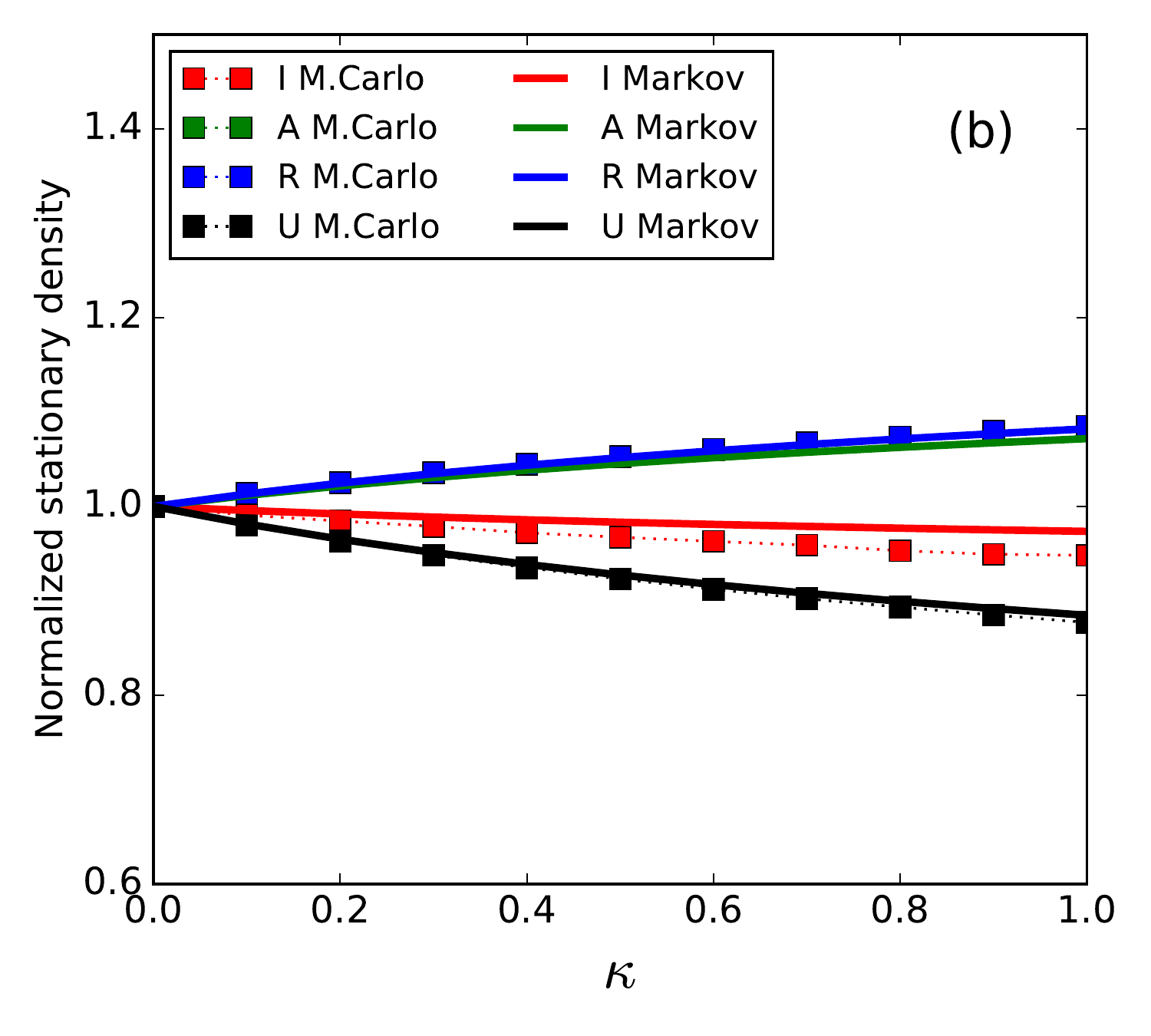}
\end{center}
\caption{Stationary densities of nodes in states I (infected), A (aware), R (stifler) and U (unaware) normalized by their values when there is no self-awareness (at $\kappa=0$) as a function of $\kappa$, for (a) $\pi = 0.1$ (slow rumor spreading) and (b) $\pi = 0.9$ (fast rumor spreading). The baseline model was used for this figure. Squares are the results of Monte Carlo simulations, whereas the solid lines are Markov chain calculations (see the appendix). The dotted lines are guides to the eyes. Other parameters are set to: $\beta = 1.0$, $\mu = 0.9$, $\gamma = 0.5$, $\alpha = 0.6$, $\Gamma = 0.0$ and $\sigma = 0.6$.}
\label{fig:stationary}
\end{figure}

To investigate in more detail how the variation of the relative timescales between the two processes affects the prevalence, we consider the behavior of the infected, aware, stifler and unaware stationary densities as a function of the parameter $\kappa$ (probability of self-awareness for an infected-unaware node). Figure \ref{fig:stationary} shows these stationary densities for two different values of $\pi$, yet for the baseline model. Each curve is normalized by its value when there is no self-awareness (i.e., $\kappa = 0$). We notice that, in both cases, the self-awareness is beneficial to the disease prevention, as the densities of aware (A) and stifler (R) nodes increase, thus reducing the density of unaware (U) nodes and the disease prevalence (I). Figure \ref{fig:stationary} also shows good agreement between Markov chain method and Monte Carlo simulations.

\begin{figure}[t]
\begin{center}
 \includegraphics[width=0.4\textwidth]{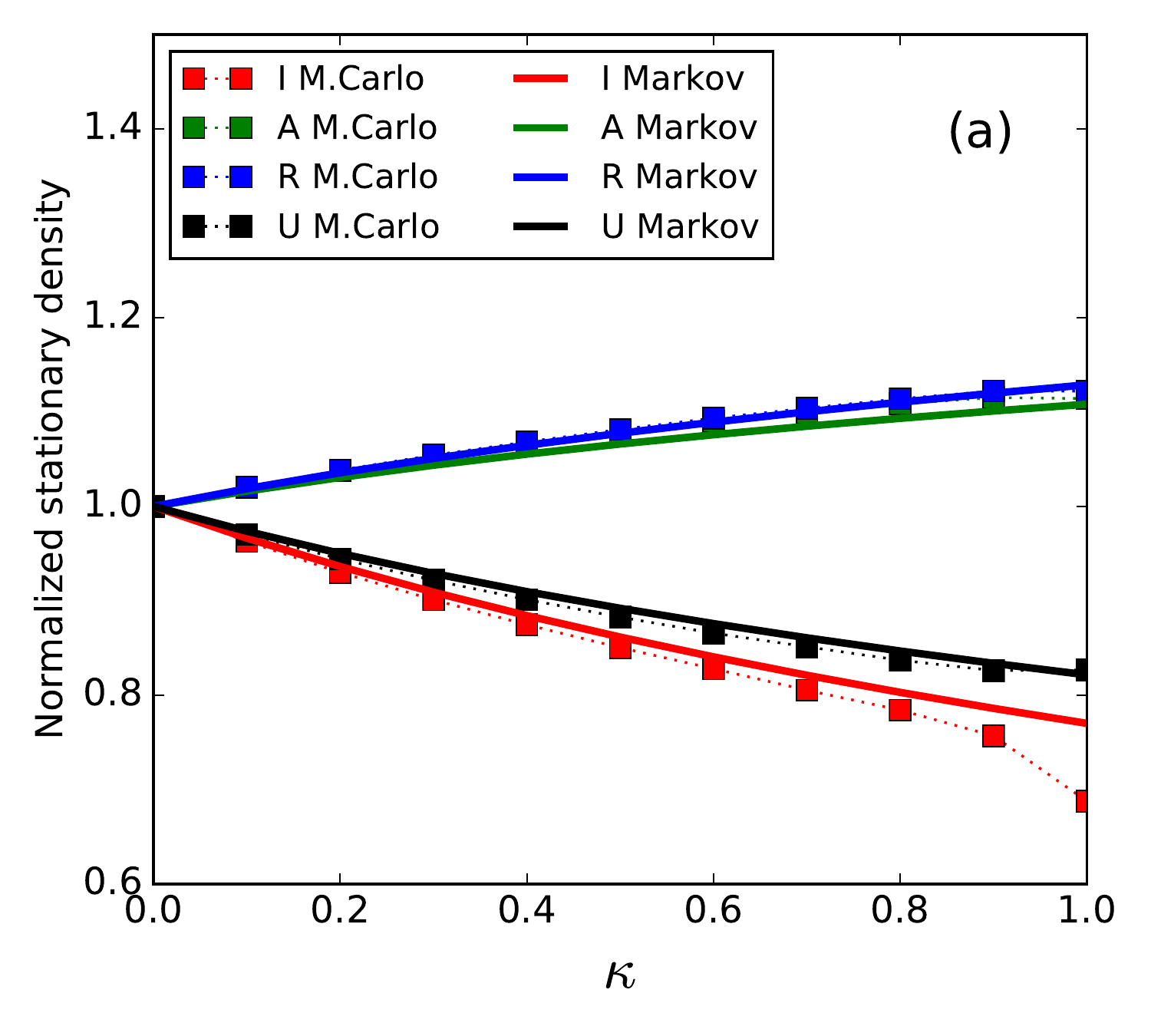}
 \includegraphics[width=.4\textwidth]{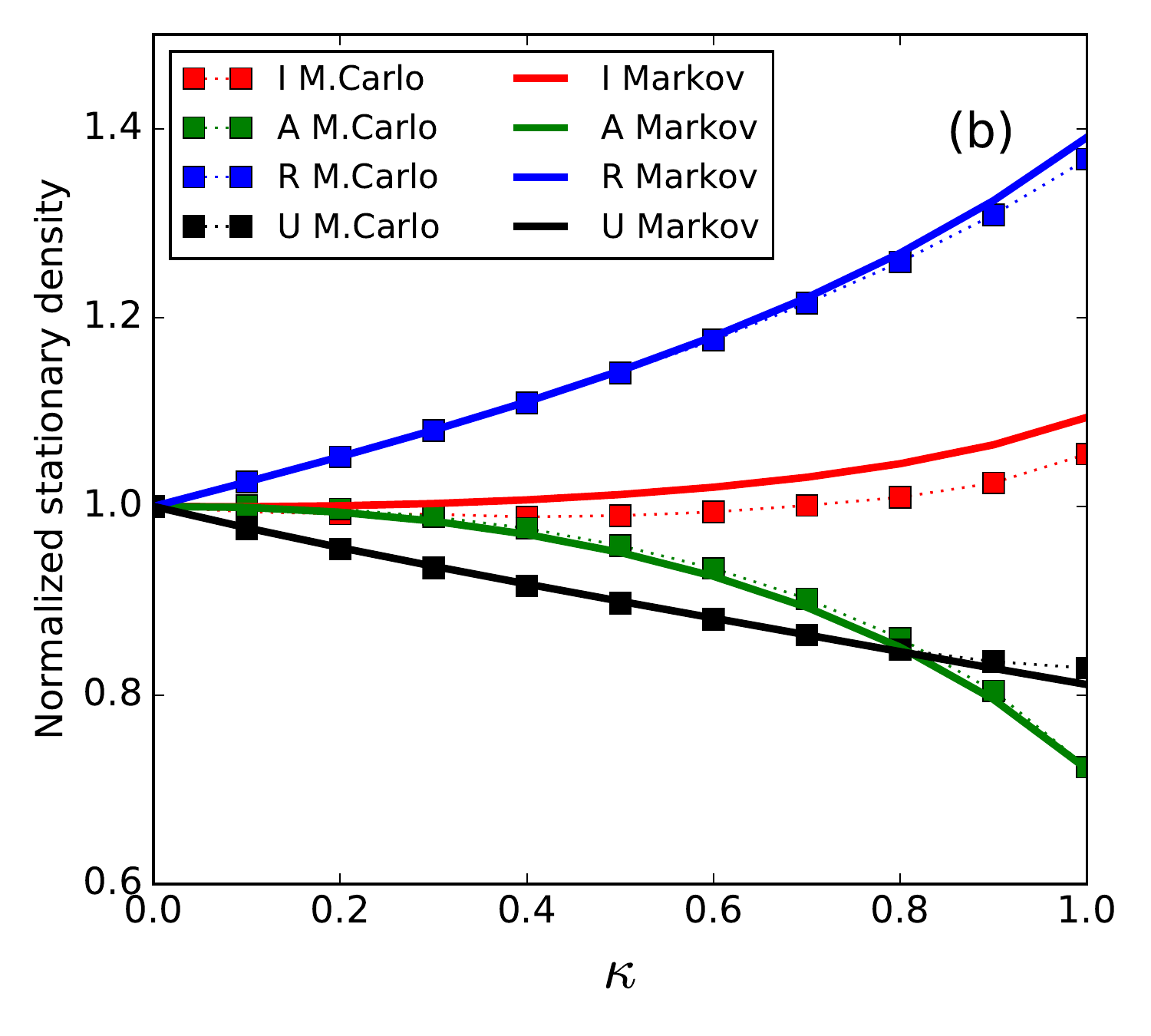}
\end{center}
\caption{Normalized stationary densities vs $\kappa$ using the modified model (see text), for (a) $\pi = 0.1$ (fast epidemic spreading) and (b) $\pi = 0.9$ (fast rumor propagation).  Squares and solid lines correspond to MC simulations and the Markov chain approach, respectively. Parameter values are the same as those in Fig.~\ref{fig:stationary}.}
\label{fig:stationary_modified}
\end{figure}

\begin{figure}[!t]
    \centering
    \includegraphics[width=0.40\textwidth]{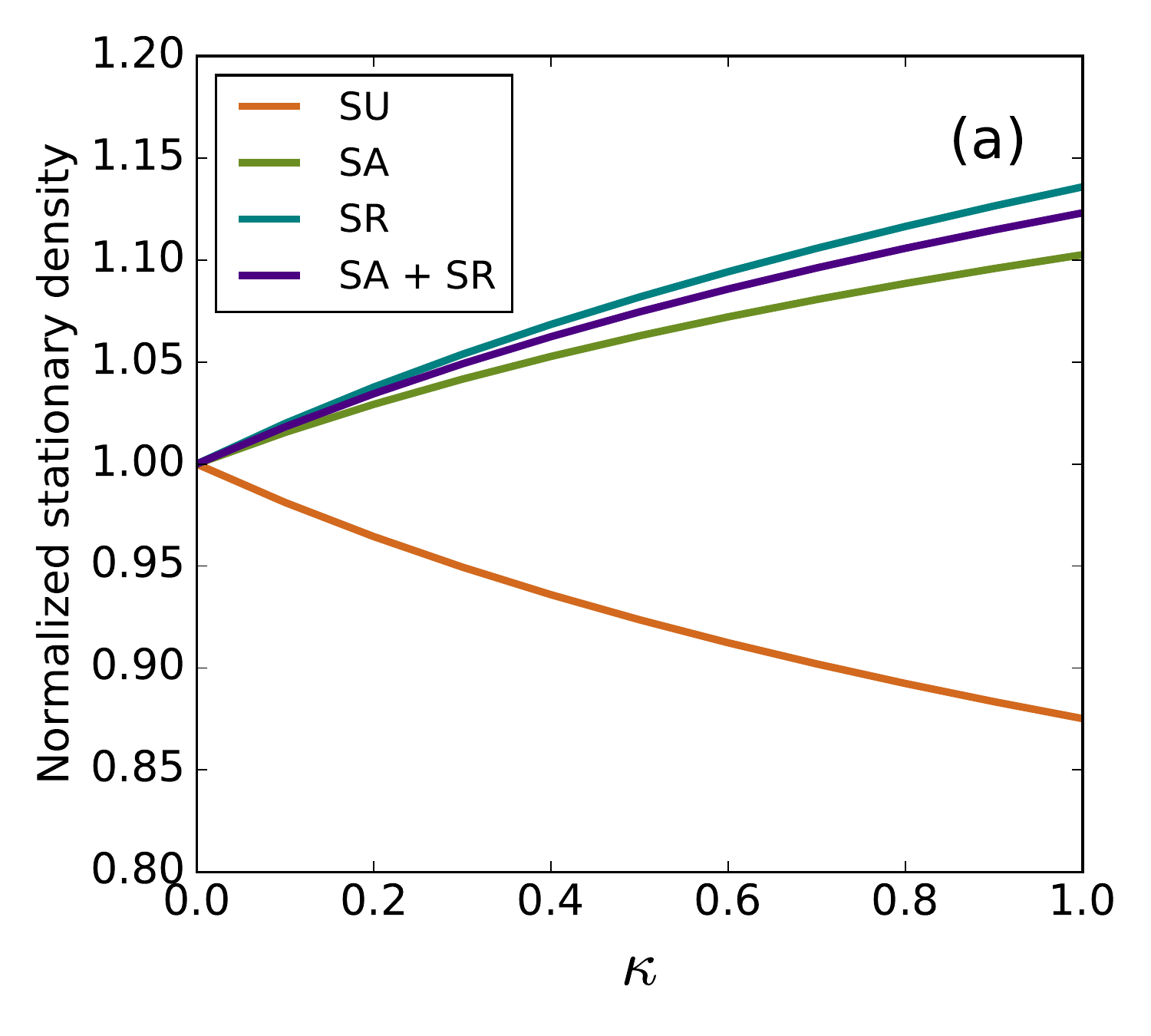}
    \includegraphics[width=0.40\textwidth]{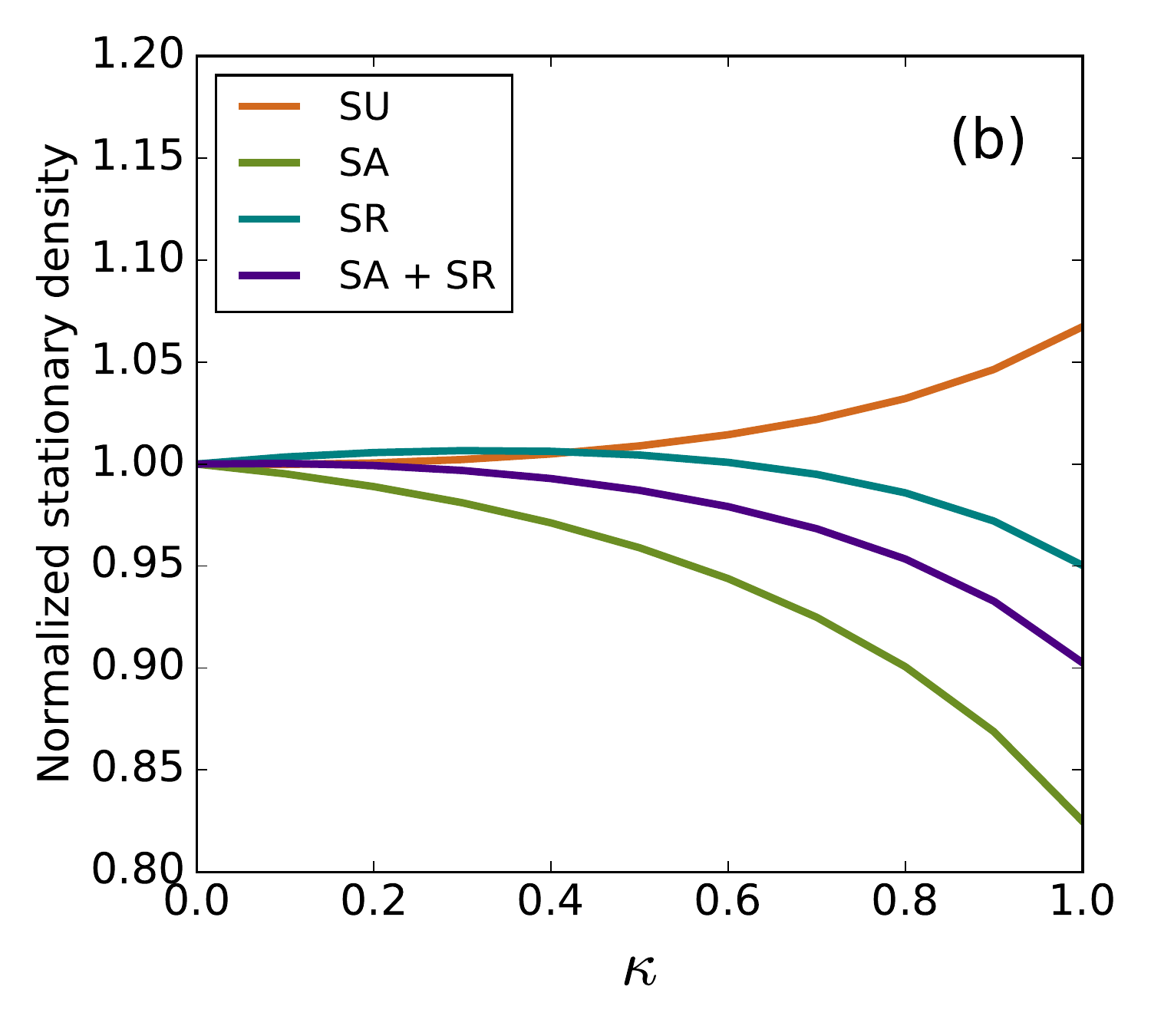}
    \caption{Stationary densities for the susceptible nodes as a function of the self-awareness probability $\kappa$, normalized by their values with $\kappa = 0$, using the modified model, for different values of the timescale parameter $\pi$. The increasing on SU population with $\kappa$ for $\pi = 0.9$ helps explaining the behavior in Fig. \ref{fig:stationary_modified}. Other parameters are set to: $\beta = 1.0$, $\mu = 0.9$, $\gamma = 0.5$, $\alpha = 0.6$, $\Gamma = 0.0$ and $\sigma = 0.6$.}
    \label{fig:twolayer_comparts}
\end{figure}

The picture changes if we consider the modified model, described in section \ref{sec:modified_model}. Figure \ref{fig:stationary_modified} shows the same plot as in figure \ref{fig:stationary} for the modified model, also with Monte Carlo and Markov chain simulations. 
For $\pi = 0.1$, as it happens in the baseline model, both densities of aware and stifler nodes increase with $\kappa$. However, for $\pi = 0.9$, when the rumor propagates faster than the disease, the fraction of aware (A) nodes decreases with $\kappa$, whereas the fraction of stiflers (R) increases very rapidly with $\kappa$.  This means that the propagation of the information is hindered by the self-awareness of infected stiflers, as they resist to forget the information (unaware) and then become aware (spreaders) again. Notice that, although the stifler population increases in comparison to the aware population, the density of unaware nodes still decreases with $\kappa$, meaning that less individuals are unprotected from the disease. Nevertheless, the prevalence (I) increases with $\kappa$ in this case. 

This is another counterintuitive result, because the disease prevalence is greater even though the unaware population is smaller. To understand this, we look at the susceptible population: if most susceptible individuals are unaware of the disease, the information is concentrated at infected nodes and thus is not effective in controlling the disease. In figure \ref{fig:twolayer_comparts}, we study the relative distribution of the susceptible population between unaware (SU), aware (SA), stifler (SR) and the combination of the previous two (SA + SR), using the modified model only. For the case of slower information ($\pi = 0.1$), the fraction of informed susceptible nodes (SA, SR) increases with $\kappa$, as expected. However, when $\pi = 0.9$, the opposite happens: the fraction of SA and SR decreases and the fraction of susceptible-unaware (SU) nodes increases, meaning 
that the fraction of susceptible nodes that are protected by the information decreases with $\kappa$ in this case.
Therefore, even if the unaware population is reduced with $\kappa$ (as reported in figure \ref{fig:stationary_modified}.b), the information 
is actually concentrated at infected individuals, making the protection inefficient. In other words, the fraction of informed individuals always increases with $\kappa$ but, for large $\pi$, susceptible individuals become less informed as $\kappa$ increases, and thus the number of infections increases.

\begin{figure}[t]
\begin{center}
\includegraphics[width=0.4\textwidth]{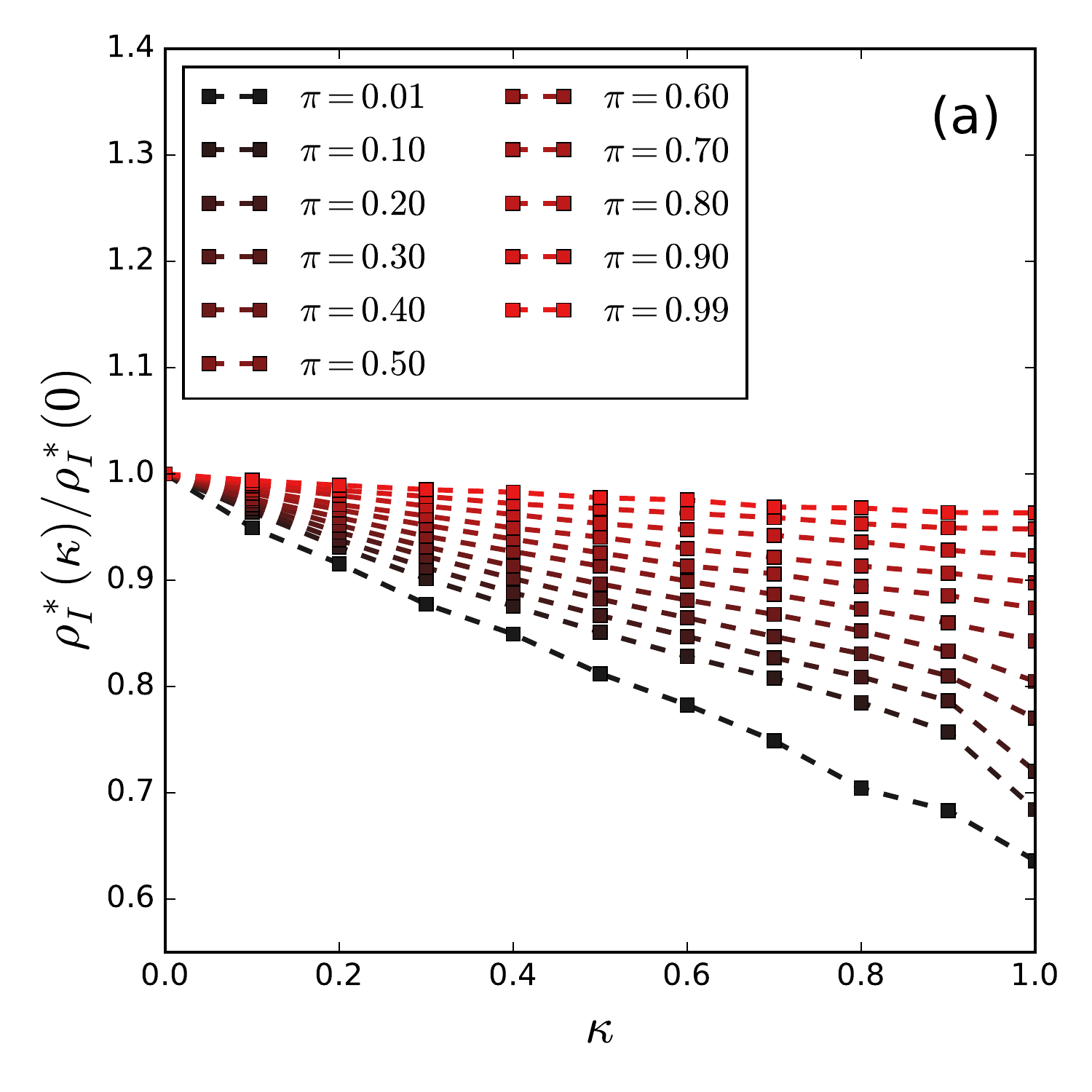}
\includegraphics[width=0.4\textwidth]{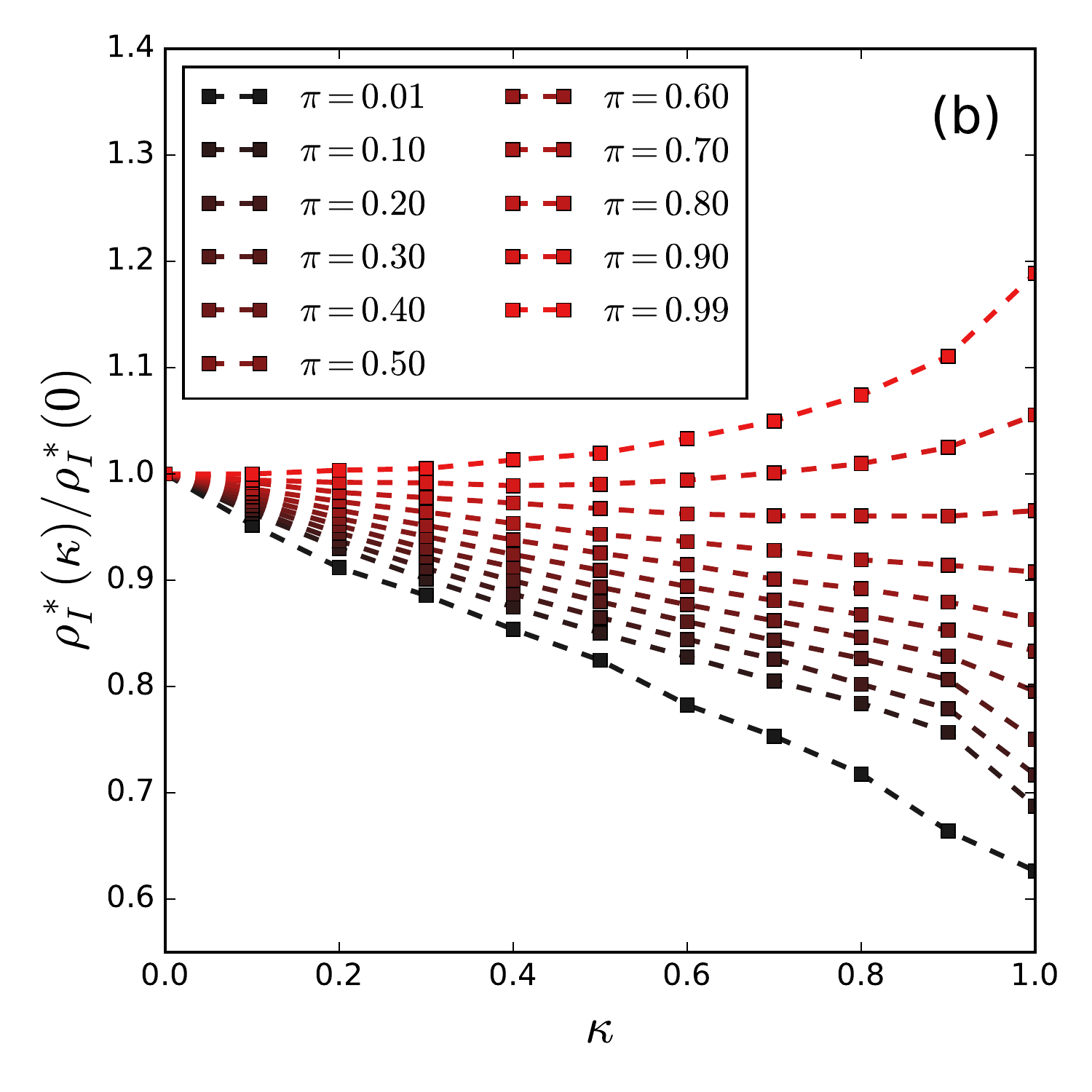}
\end{center}
\caption{Normalized disease prevalence $\rho_I^*(\kappa)/\rho_I^*(\kappa=0)$ vs $\kappa$ for the (a) baseline and (b) modified models. The values of the timescale parameter $\pi$ increase from the darker to the brighter color, showing how the curves change their behavior with $\kappa$ as $\pi$ increases. Other parameters are set to: $\beta = 1.0$, $\mu = 0.9$, $\gamma = 0.5$, $\alpha = 0.6$, $\Gamma = 0.0$, $\sigma = 0.6$.}
\label{fig:weirdkappa_sf}
\end{figure}

Hence, we conclude that the timescale, controlled by the parameter $\pi$, plays a fundamental role on the prevalence, meaning that the relative timescale between epidemics and information determines if the self-awareness is beneficial or not for the disease prevention. We also study how the parameter $\pi$ changes the behavior of the prevalence with $\kappa$ on the modified model, by analyzing the prevalence $\rho_I^*$ vs $\kappa$ curves for eleven different values of $\pi$. Figure \ref{fig:weirdkappa_sf} shows such curves, normalized by the value of the prevalence when $\kappa = 0$.

By analyzing the plots in figure \ref{fig:weirdkappa_sf}, we can conceive the influence of the timescale. For small $\pi$ (faster epidemics, slower information), the prevalence exhibits its normal decreasing behavior with $\kappa$ for both baseline and modified models. On the other hand, for larger $\pi$ (slower epidemics, faster information), the curves for the modified model flip their slope for larger $\kappa$ values, whereas they maintain the same behavior for the baseline model. This means that, when the informational processes are considerably faster than the disease transmission, the self-awareness process can generate too many stiflers and impair the information spreading, increasing the prevalence. For both baseline and modified models, the timescale plays an important role on determining the effectiveness of the information on reducing the disease prevalence.

The results presented so far were taken using a pair of scale-free (SF) networks. One natural question is whether the observed phenomena are due to the particular topology that we used. To answer that, we simulated the (modified) model using combinations of two other topologies: the Watts-Strogatz (WS)\cite{watts1998collective} and Erd\H{o}s-Rényi (ER)\cite{Erdos060} models. The ER layers were generated with connection probability $p = 0.008$, which produces an average degree of $\mean{k} \approx 8$ (for $N = 1000$ nodes). The WS layers were generated with average degree $\mean{k} = 8$ and rewiring probability $p_r = 0.01$, which produces layers with average clustering coefficient $C \approx 0.47$ and average shortest path length $l \approx 5$. For the simulations, we used the following pairs of epidemic/informational (in this order) layers: ER/ER, ER/SF, ER/WS, WS/WS, SF/WS and the previously used SF/SF pair.

\begin{figure}[!h]
    \centering
    \includegraphics[width=0.42\textwidth]{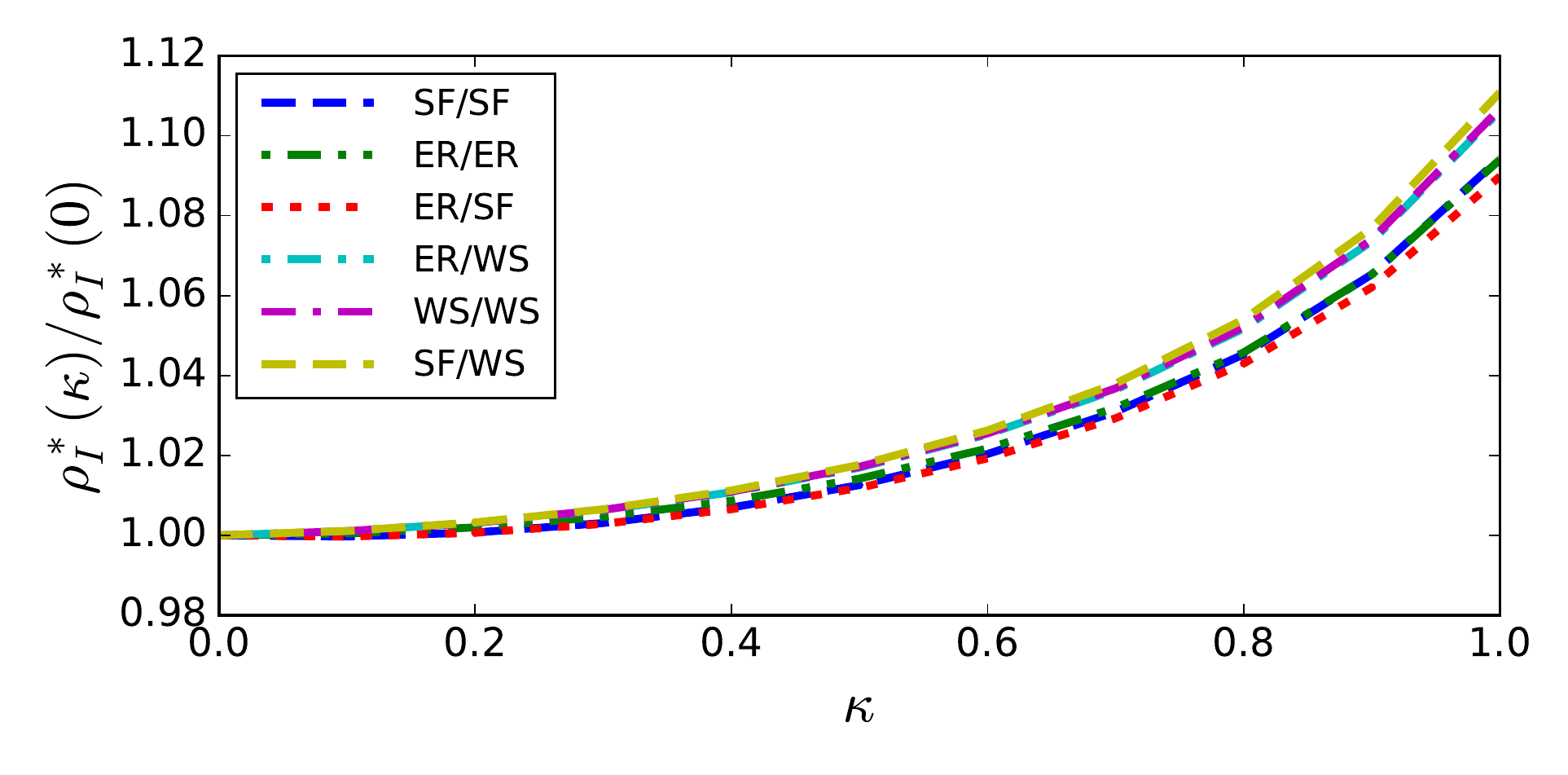}
    \includegraphics[width=0.42\textwidth]{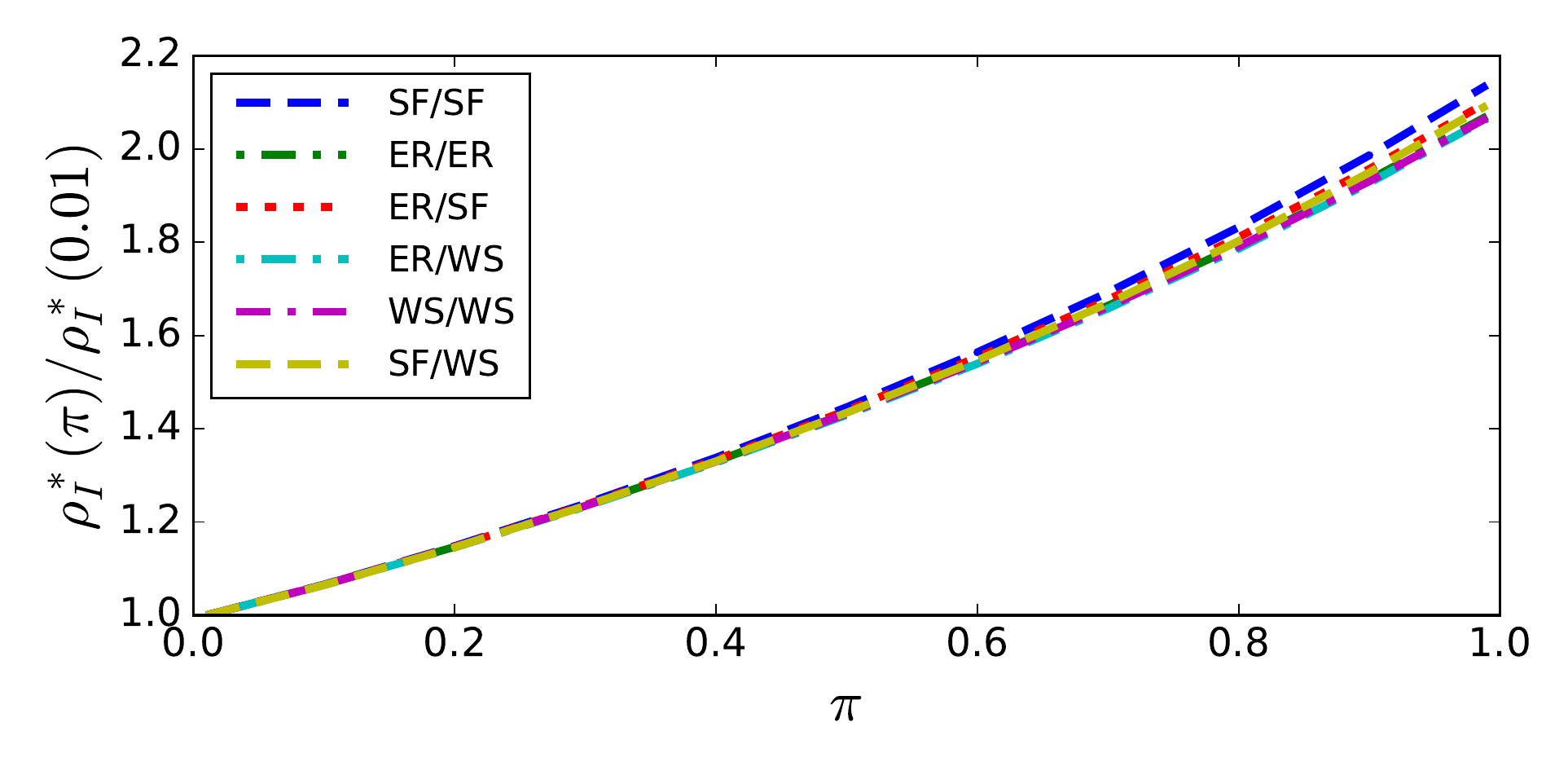}
    \caption{Markov chain calculations of the normalized prevalence for the modified model, using different pairs of network models between the configurational scale-free (SF), Erd\H{o}s-Rényi (ER) and Watts-Strogatz (WS) models. On the upper plot, the prevalence is shown as a function of the self-awareness parameter ($\kappa$) for a fixed value of $\pi = 0.9$, and on the lower plot it is shown as a function of $\pi$ for a fixed value of $\kappa = 0.8$. Other parameters are set to the same values as in figure \ref{fig:weirdkappa_sf}.}
    \label{fig:other_topologies}
\end{figure}

Figure \ref{fig:other_topologies} shows the results of Markov chain calculations using other topologies. On the upper plot, we show the prevalence as a function of the self-awareness parameter $\kappa$, for a high value of the timescale ($\pi = 0.9$, meaning faster informational processes). On the lower plot, we show the prevalence as a function of the timescale parameter $\pi$ for a fixed value ($\kappa = 0.8$) of the self-awareness. All prevalence values are normalized by the first value of the sequence. For all pairs of topologies, the basic results that we presented before - the increasing of the prevalence with $\pi$ and the reversed behavior with $\kappa$ for high values of $\pi$ are consistently preserved.

\section{Epidemic critical point and phase diagrams}

Following the procedure proposed in \cite{gomez_2010}, we can calculate the epidemic critical point for our model. We show in appendix \ref{sec:epid_threshold} that the phase transition curve between the endemic and the healthy state is, for both baseline and modified models, given by $\beta / \mu = (\Lambda_{\text{max}}(H))^{-1}$, where the elements of matrix $H$ are defined as:

\begin{equation}
    H_{ij} =  [1 - (p_A^i + p_R^i)(1 - \Gamma)] A_{ij}
\end{equation}

Where $A_{ij}$ is the epidemic layer adjacency matrix and $\Lambda_{\text{max}}$ represents the greatest eigenvalue. This result is the same as in the model from Granell with no mass media \cite{arenas_2013}, only replacing the probability that node $i$ is simply aware $p_A^i$ by the probability that it is ``protected'' $p_A^i + p_R^i$, whose value is calculated by solving the awareness equations without epidemics.

One first notorious fact is that the epidemic critical point does not depend on the relative timescale $\pi$, as it does not change the individual ``forces'' of the epidemic and informational processes. It also does not depend on $\kappa$, as self awareness is irrelevant when the prevalence is very small.

\begin{figure}[h]
    \centering
    \setbox1=\hbox{\includegraphics[height=2.98cm]{example-image-b}} 
    \includegraphics[width=0.499\textwidth]{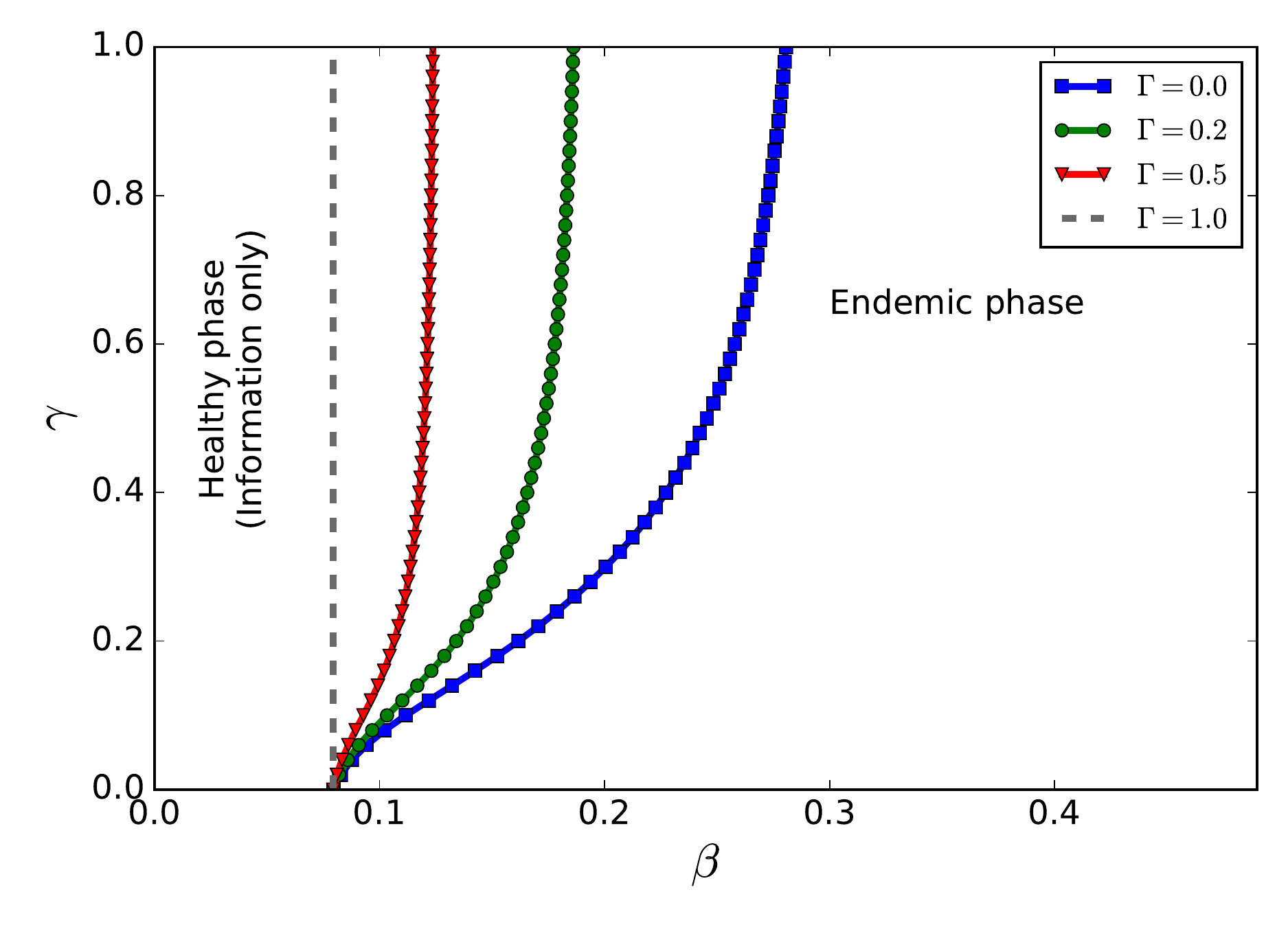}\llap{\makebox[\wd1][l]{\raisebox{0.97cm}{\includegraphics[height=2.7cm]{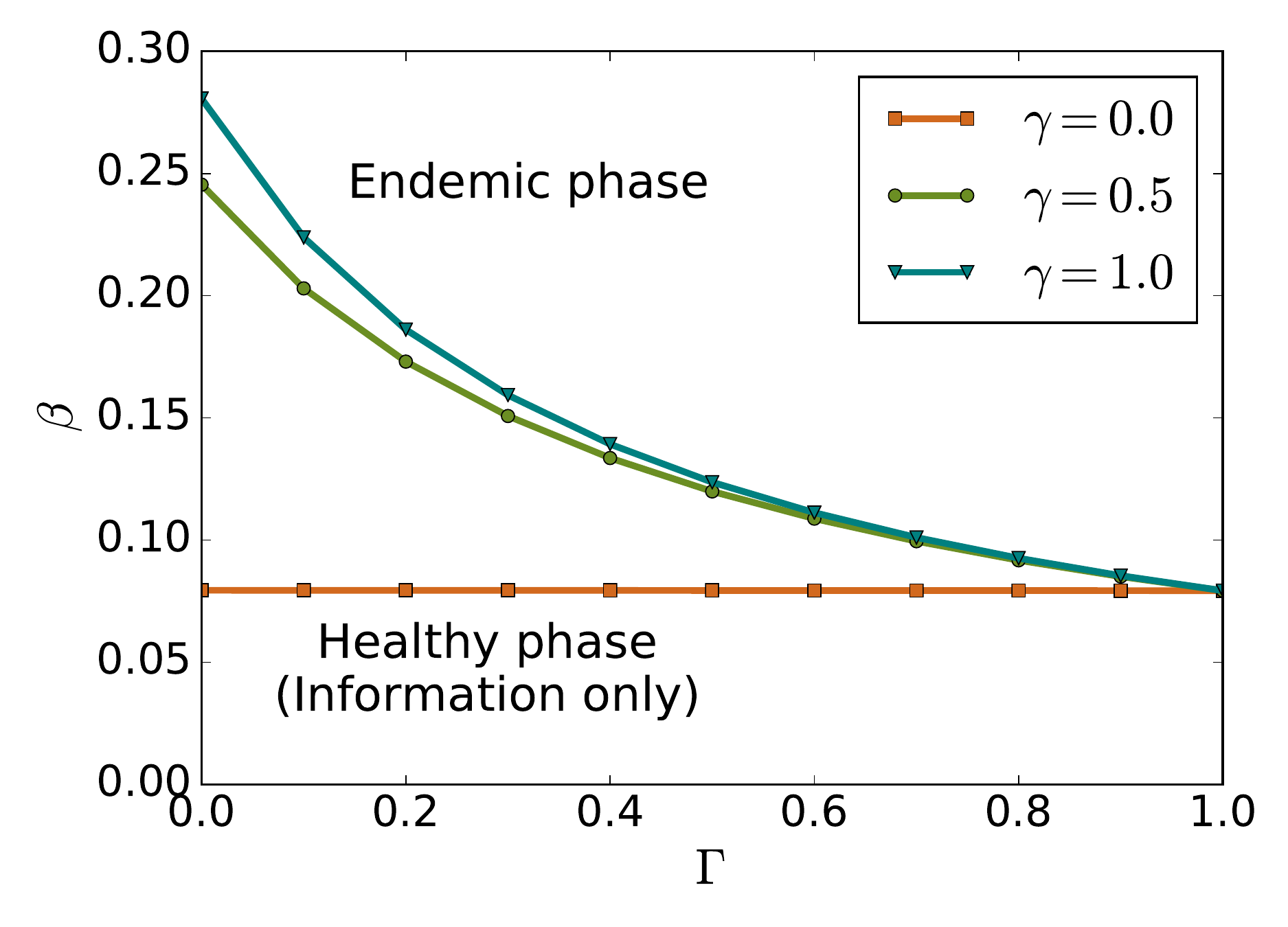}}}}
    \caption{Phase diagram of the model, showing the phase transition curve between the healthy and endemic phases for different values of $\Gamma$. The inset shows the critical epidemic transmission probability $\beta$ as a function of the protection factor $\Gamma$, for different values of the information transmission probability $\gamma$. The diagram is the same for the modified and baseline models, and does not depend on $\pi$ and $\kappa$. Other parameters are set to: $\mu = 0.9$, $\alpha = 0.6$, $\sigma = 0.6$.}
    \label{fig:phase_diagrams}
\end{figure}

Figure \ref{fig:phase_diagrams} shows the phase transition curves in the $\beta$ x $\gamma$ plane, for four different values of the protection factor $\Gamma$. At the left of each curve lies the healthy phase (no disease in stationary state), whereas the endemic phase ($\rho_I^* > 0$) is at the right. On the inset, we show how the epidemic critical point depends on the protection factor $\Gamma$.

One of the main differences to the simpler SIS/UAU model presented by Granell and others\cite{arenas_2013} is that the ``metacritical'' point is not present, thus resembling the similar model with mass media presented in \cite{arenas_2014} by the same authors. This happens because the rumor model UARU has no phase transition, i.e., there is always a fraction of nodes that is aware. Therefore, our SIS/UARU model presents only two phases: healthy and endemic, lacking a phase in which both disease and information are extinct.

In our model, therefore, the population can be either informed and healthy or informed and endemic. Although information exists even without disease, the density of informed individuals is enhanced by the presence of disease (provided that there is self-awareness), as it can be seen in figures \ref{fig:stationary} and \ref{fig:stationary_modified}.

\section{Conclusions}

We have analyzed the effect of information awareness to prevent the transmission of disease in multiplex networks. We have considered the Maki-Thompson rumor model for the propagation of the information, which incorporates a forgetting mechanism not included in previous related models.  Besides, the rumor and disease spread at the same time but under different timescales that control the relative speeds of these two processes. We have verified that the information helps to reduce the prevalence and increase the epidemic threshold of the disease. We have also observed that self-awareness, which keeps infected individuals aware of their condition, is a very effective mechanism for reducing the disease prevalence.  However, in the case that the information spreads much faster than the disease, large values of self-awareness can lead to the counterintuitive result of a higher prevalence.  This happens because the self-awareness can generate such an excessive number of stiflers that impair the propagation of information, with the overall effect of increasing the prevalence.  Therefore, the relative timescales between the information and infection processes determines whether the information awareness is beneficial or not for the magnitude of the epidemics.  In this way, our work highlights the important role played by infected individuals who help spreading the information about the disease, reducing the disease transmission and the outbreak.

Although our results are obtained only by numerical simulations on multiplex networks, we show that the results are robust with respect to the topology, suggesting that they can be extended even for homogeneously mixed populations. The topology, therefore, may generate quantitative effects, but not change the qualitative behavior of the model. We finally investigated the epidemic critical behavior, comparing it to previous models on the literature. Although the relative timescale (controlled by $\pi$) has an important influence on the disease prevalence and its behavior with the model parameters, it does not affect the epidemic threshold.

Our investigations can also be extended by considering other dynamics for rumor and disease spreading, as well as networks presenting assortativity and community organization. As a general conclusion, this work provides a motivation for studying other interacting processes using flexible timescales. It could be of great value for the community to understand when the results and critical behavior of dynamical processes are affected or not by timescale differences between each process.

\acknowledgments
PCVS thanks FAPESP for the PhD grant 2016/24555-0.
Research carried out using the computational resources of the Center for Mathematical Sciences Applied to Industry (CeMEAI) funded by FAPESP (grant 2013/07375-0). FAR acknowledges CNPq (grant 307974/2013-8) and FAPESP (grants 17/50144-0 and 16/25682-5) for the financial support given for his research. FAR gratefully acknowledges support from The Leverhulme Trust for the Visiting Professorship provided. This researched is also supported by FAPESP (grant 2015/50122-0) and DFG-GRTK (grant 1740/2). YM acknowledges partial support from the Government of Arag\'on, Spain through grant E36-17R, and by MINECO and FEDER funds (grant FIS2017-87519-P). FV acknowledges financial support from CONICET (PIP 0443/2014) and from Agencia Nacional de Promoción Cienítfica y Tecnológica (PICT 2016 Nro 201-0215). Y.M. also acknowledges partial support from Intesa Sanpaolo Innovation Center. The funders had no role in study design, data collection, and analysis, decision to publish, or preparation of the manuscript.

\appendix
\section{Markov chain approach for the SIS/UARU model}

\label{sec:markov_chain}

In order to predict the behavior of the model in a double-layer network, we develop a microscopic Markov chain approach to write dynamical equations for relevant probabilities of our system. For that purpose, we follow the methodology described in \cite{gomez_2010}.

For each node $i$ of the network and for each time stamp $t$, the probabilities that it is in each possible state of the model is defined as $p$. Such state can either be from a single process (e.g., $p_I^i(t)$ is the probability that node $i$ is infected (I) at time $t$) or a ``composite'' state (e.g., $p_{SU}^i(t)$, $p_{IA}^i(t)$, etc.). For convenience, we label the corresponding nodes in different layers with the same number.

The first step is to build the transition trees for all possible changes of states and their respective transition probabilities. For each tree, we represent the root as one of the possible composite states of a node (SU, SA, SR, IU, IA and IR) at time $t$, and the leaves at each of the possible resulting states at time $t+1$, starting from the state at time $t$. The branches represent each of the possible transitions. The probabilities of such transitions are written above the corresponding branches.

\begin{figure}[h]
    \centering
    \includegraphics[width=0.47\textwidth]{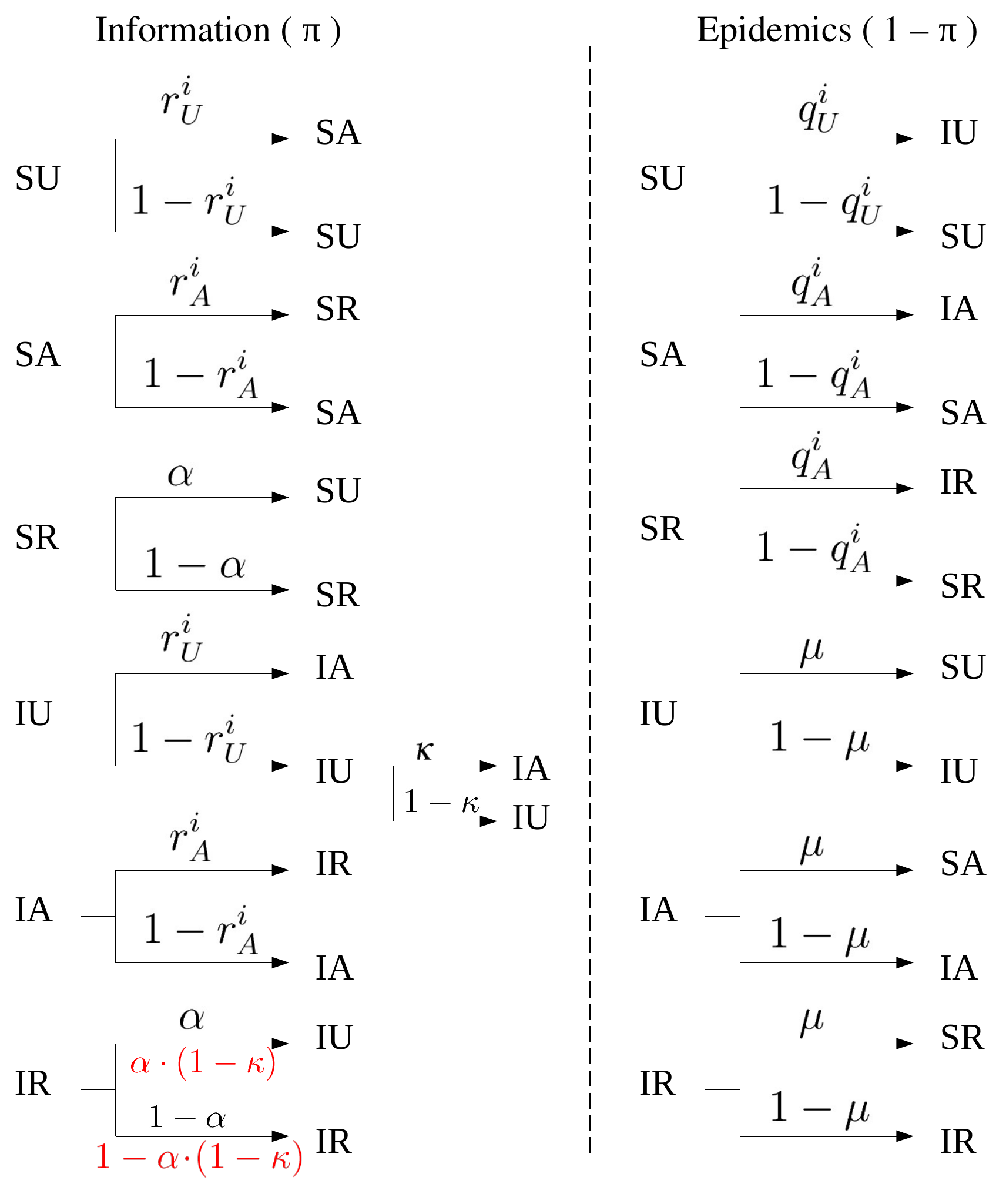}
    \caption{Probability trees with all the possible transitions for each state. The informational group has an associated probability of $\pi$, whereas the epidemic group carries the complementary probability $1 - \pi$. For the modified model, the $\text{IR} \rightarrow \text{IU}$ transition follows the factors in red (instead of the ones in black). 
    }
    \label{fig:probab_trees_sisuaru}
\end{figure}

As described previously for the numerical simulations, we separate the transitions into two groups - the epidemic and informational - and only one of the transitions groups is performed in a time step. Figure \ref{fig:probab_trees_sisuaru} shows the trees for the SIS/UARU model. The baseline model corresponds to all the factors in black. The modified model has the same factors of the baseline model, except for the $\text{IR} \rightarrow \text{IU}$ transition, in which the correct factors are displayed in red. The modification is interpreted as a reduction on the forgetting probability for IR nodes.

The probability of each event on the informational side of figure \ref{fig:probab_trees_sisuaru} is multiplied by $\pi$, which is the probability that the informational group is chosen to be updated in the current time step. On the other hand, probabilities from the epidemic side carry a factor of $1 - \pi$. Therefore, for instance, the probability that an infected-aware (IA) node gets healed and becomes susceptible-aware (SA) is of $(1-\pi)\cdot\mu$, following the corresponding probability tree on the epidemic group. 

The transition probabilities for processes which involve contact with neighboring nodes, namely $q_U^i$ (infection of an unaware node), $q_A^i$ (infection of an aware node), $r_U^i$ (awareness by contacting an aware neighbor) and $r_A^i$ (``stifling'' - lost of interest) are defined by the following set of equations:
\begin{eqnarray}
\label{eq:trans_prob_qu}
    q_U^i &=& 1 - \prod_{j}(1 - A_{ij} \, p_I^j \, \beta), \\
    q_A^i &=& 1 - \prod_{j}(1 - A_{ij} \, p_I^j \, \Gamma\beta),  \\
    r_U^i &=& 1 - \prod_{j}(1 - B_{ij} \, p_A^j \, \gamma), \\
\label{eq:trans_prob_ra}
    r_A^i &=& 1 - \prod_{j}(1 - B_{ij} \, (p_A^j + p_R^j) \, \sigma )
\end{eqnarray}

where $A_{ij}$ and $B_{ij}$ represent the adjacency matrices for the epidemic and informational layers, respectively. Here, we point out that our goal is to study the stationary state of the system, in which all probabilities do not change in time. Therefore, the time label $t$ of all probabilities defined here (e.g, $p_{SU}^i(t), r_{U}^i(t)$) were removed. 

Based on the transition trees drawn in Fig. \ref{fig:probab_trees_sisuaru}, we can write down the Markov chain equations for the probabilities of each node $i$ being in each of the six compartments (SU, SA, SR, IU, IA, IR) of the model as a fixed point set of equations, in which the time dependence is already removed: 

\begin{eqnarray}
\label{eq:markov_chains_su} 
    \nonumber p_{SU}^i &=& p_{SU}^i [\pi  (1 - r_U^i) + (1-\pi)(1 - q_U^i)] + \\ 
    \nonumber    &\quad +& p_{SR}^i [\pi \alpha] + \\
                 &\quad +& p_{IU}^i [(1-\pi) \mu] \\
\label{eq:markov_chains_sa} 
    \nonumber p_{SA}^i &=& p_{SU}^i [\pi r_U^i] + \\
    \nonumber    &\quad +& p_{SA}^i [\pi(1-r_A^i) + (1-\pi) (1 - q_ A^i)] + \\
                 &\quad +& p_{IA}^i [(1-\pi) \mu] \\
\label{eq:markov_chains_sr} 
    \nonumber p_{SR}^i &=& p_{SA}^i [\pi r_A^i] + \\
    \nonumber    &\quad +& p_{SR}^i [\pi(1-\alpha) + (1-\pi) (1 - q_A^i)] + \\
                 &\quad +& p_{IR}^i [(1-\pi) \mu] \\
\label{eq:markov_chains_iu} 
    \nonumber p_{IU}^i &=& p_{SU}^i [(1-\pi) q_U^i] + \\
    \nonumber    &\quad +& p_{IU}^i [\pi(1-r_U^i)(1-\kappa) + (1-\pi) (1 - \mu)] + \\
                 &\quad +& p_{IR}^i [\pi \alpha (1 - \kappa)] \\
\label{eq:markov_chains_ia} 
    \nonumber p_{IA}^i &=& p_{SA}^i [(1-\pi) q_A^i] + \\
    \nonumber    &\quad +& p_{IU}^i [\pi (r_U^i + (1-r_U^i) \kappa) ] + \\
                 &\quad +& p_{IA}^i [\pi  (1 - r_A^i) + (1-\pi) (1-\mu)] \\
\label{eq:markov_chains_ir}
    \nonumber p_{IR}^i &=& p_{SR}^i [(1-\pi) q_A^i] + \\*
    \nonumber    &\quad +& p_{IA}^i [\pi r_A^i] + \\
                 &\quad +& p_{IR}^i [\pi  (\alpha \kappa + 1 - \alpha) + (1-\pi) (1-\mu)] \\
    \nonumber
\end{eqnarray}

Equations \ref{eq:markov_chains_sa} to \ref{eq:markov_chains_ir} represent the baseline model. For the modified model, in which the IR ${\rightarrow}$ IU has a modified probability, equations \ref{eq:markov_chains_iu} and \ref{eq:markov_chains_ir} are respectively replaced by:

\begin{eqnarray}
\label{eq:modif_markov_chains_iu} 
    \nonumber p_{IU}^i &=& p_{SU}^i [(1-\pi) q_U^i] + \\
    \nonumber    &\quad +& p_{IU}^i [\pi(1-r_U^i)(1-\kappa) + (1-\pi) (1 - \mu)] + \\
                 &\quad +& p_{IR}^i [\pi \alpha] \\
\label{eq:modif_markov_chains_ir}
    \nonumber p_{IR}^i &=& p_{SR}^i [(1-\pi) q_A^i] + \\*
    \nonumber    &\quad +& p_{IA}^i [\pi r_A^i] + \\
                 &\quad +& p_{IR}^i [\pi  (1 - \alpha) + (1-\pi) (1-\mu)] \\
    \nonumber
\end{eqnarray}

We solve the system of $6N$ equations (where $N$ is the number of nodes on the network) by the fixed point method, in which the LHS values are updated by applying previous values at the RHS expressions. As explained in the main text, the initial conditions are set to: $p_{IA}^i = 0.2$, $p_{SU}^i = 0.8$ and $p_{IU}^i = p_{IR}^i = p_{SA}^i = p_{SR}^i = 0$, for $i = 0, 1, ..., N-1$. The solutions of these equations are shown in figures \ref{fig:rhoi_vs_beta}, \ref{fig:stationary} and \ref{fig:stationary_modified} as solid lines, where we can see a good agreement between the Markov chain predictions and Monte Carlo simulations.

\section{Epidemic critical point}
\label{sec:epid_threshold}

From the Markov chain equations, we can derive the epidemic critical point between the healthy and the endemic phases. The basic idea is to analyze the stability of the healthy solution $p_I^i = 0$, $i = 1, 2, ..., N$, using a perturbative approach.

We first add equations \ref{eq:markov_chains_iu} to \ref{eq:markov_chains_ir} to obtain the evolution of the probability $p_{I}^i = p_{IU}^i + p_{IA}^i + p_{IR}^i$ that node $i$ is infected:

\begin{eqnarray}
    \label{eq:p_i_markov}
    p_I^i =& \;(1 - \pi) [p_I^i (1 - \mu) + p_{SU}^i q_U^i + \nonumber \\ 
    & + \; (p_{SA}^i + p_{SR}^i)  q_A^i  ] \; + \;  \pi p_I^i
\end{eqnarray}

Notice that this equation holds both for the baseline and modified models, as the informational terms add up to $\pi p_ I^i$ in any case. We now use the following approximation for the $q_U^i$ and $q_A^i$ transition probabilities, which is valid when $p_I^i$ is sufficiently small for any node $i$:

\begin{eqnarray}
    \label{eq:qUi_approx}
    q_U^i \approx \beta \sum_j A_{ij} p_I^j \\
    \label{eq:qAi_approx}
    q_A^i \approx \Gamma \beta \sum_j A_{ij} p_I^j
\end{eqnarray}

Rewriting equation \ref{eq:p_i_markov} with these approximations yields:

\begin{align}
    p_I^i \approx & (1 - \pi) \Big{\{}p_I^i (1 - \mu) +\beta\;[ p_{SU}^i + \nonumber \\ 
    & + \; \Gamma(p_{SA}^i + p_{SR}^i) ] \sum_j A_{ij} p_I^j \Big{\}} \; + \;  \pi p_I^i
\end{align}

Sending the terms with $p_I^i$ to LHS and leaving the terms with $\sum_j A_{ij} p_I^i$ on the RHS, we end up with the following self-consistent relation for $p_I^i$, which does not depend on the timescale $\pi$:

\begin{align}
    \label{eq:linearized_pIi}
    p_I^i \approx \frac{\beta}{\mu} [p_{SU}^i + \Gamma(p_{SA}^i + p_{SR}^i) ] \sum_j A_{ij} p_I^j
\end{align}

Equation \ref{eq:linearized_pIi} is a matrix equation of the shape $\overrightarrow{p} = (\beta/\mu) H \overrightarrow{p}$. The trivial solution $p_I^i = 0$ for every node $i$ is stable on equation \ref{eq:linearized_pIi} if all eigenvalues of the matrix $H$, with elements defined as:

\begin{align}
    \nonumber H_{ij} = & [p_U^i + \Gamma (p_A^i + p_R^i)] A_{ij} = \\ 
    = & [1 - (p_A^i + p_R^i)(1 - \Gamma)] A_{ij}
\end{align}

Are not greater than $\mu / \beta$. Noticed that we also approximated $p_{U}^i = p_{SU}^i + p_{IU}^i \approx p_{SU}^i$, and the same for A and R compartments. Therefore, the expression for the healthy/endemic phase transition curve is:

\begin{equation}
    \frac{\beta}{\mu} = \frac{1}{\Lambda_{\text{max}}(H)}
\end{equation}

The values of $p_A^i$ and $p_R^i$ can be found by solving the Markov chain equations for the informational model only, with no interference of the disease.

\bibliographystyle{apsrev4-1}
\bibliography{bibliography.bib}

\end{document}